\documentclass{article}
\pdfoutput=1

\PassOptionsToPackage{numbers,sort&compress}{natbib}

\usepackage[final]{nips_2018}
\usepackage[utf8]{inputenc}
\usepackage{natbib}
\usepackage{url}
\usepackage{amsmath}
\usepackage{amssymb}
\usepackage{bm}
\usepackage{venndiagram}
\usepackage[probability,notions,events,operators,sets]{cryptocode}
\usepackage{dashbox}
\usepackage{xspace}
\usepackage{hyperref}

\definecolor{turquoise}{rgb}{0.19, 0.84, 0.78}
\definecolor{prettyblue}{rgb}{0.0, 0.5, 0.69}
\definecolor{jade}{rgb}{0.0, 0.66, 0.42}

\newif\ifcomment
\commentfalse

\ifcomment
\newcommand{\seda}[1]{\marginpar{\color{purple}{sg: #1}}}
\newcommand{\bekah}[1]{\marginpar{\color{prettyblue}{ro: #1}}}
\newcommand{\bogdan}[1]{\marginpar{\color{jade}{bk: #1}}}
\newcommand{\ero}[1]{\marginpar{\color{brown}{eb: #1}}}
\newcommand{\mela}[1]{\marginpar{\color{violet}{ct: #1}}}
\else
\newcommand{\seda}[1]{}
\newcommand{\bekah}[1]{}
\newcommand{\bogdan}[1]{}
\newcommand{\ero}[1]{}
\newcommand{\mela}[1]{}
\fi

\newcommand{\newterm}[1]{\textbf{#1}}

\newcommand{\Pokemon}{Pok\'emon\xspace}

\newcommand{\deployer}{OSP\xspace}
\newcommand{\deployers}{OSPs\xspace}

\newcommand{\icurr}[1][t]{{_{i,t}}}
\newcommand{\ifuture}[1][t]{{_{i,t+1}}}

\definecolor{turquoise}{rgb}{0.19, 0.84, 0.78}
\definecolor{prettyblue}{rgb}{0.0, 0.5, 0.69}
\definecolor{jade}{rgb}{0.0, 0.66, 0.42}

\title{Questioning the assumptions behind fairness solutions\footnotemark}

\author{
	Rebekah Overdorf \\
	EPFL\\
	\texttt{rebekah.overdorf@epfl.ch} \\
	\And
	Bogdan Kulynych\\
	EPFL SPRING Lab \\
	\texttt{bogdan.kulynych@epfl.ch} \\
	\And
	Ero Balsa\\
	imec-COSIC KU Leuven\\
	ebalsa@esat.kuleuven.be\\
	\And
	Carmela Troncoso \\
	EPFL SPRING Lab \\
	carmela.troncoso@epfl.ch \\
	\And
	Seda G\"urses \\
	imec-COSIC KU Leuven \\
	sguerses@esat.kuleuven.be \\
}

\begin{document}
\maketitle
\renewcommand{\thefootnote}{\fnsymbol{footnote}}
\footnotetext{This position paper is based on the full paper \cite{pots} by the same authors.}

\begin{abstract}


In addition to their benefits, optimization systems can have negative economic, moral, social, and political effects on populations as well as their environments. Frameworks like fairness have been proposed to aid service providers in addressing subsequent bias and discrimination during data collection and algorithm design. However, recent reports of neglect, unresponsiveness, and malevolence cast doubt on whether service providers can effectively implement fairness solutions. 
These reports invite us to revisit assumptions made about the service providers in fairness solutions. Namely, that service providers have (i) the incentives or (ii) the means to mitigate optimization externalities. Moreover, the environmental impact of these systems suggests that we need (iii) novel frameworks that consider systems other than algorithmic decision-making and recommender systems, 
and (iv) solutions that go beyond removing related algorithmic biases. Going forward, we propose Protective Optimization Technologies that enable optimization subjects to defend against negative consequences of optimization systems. 



\end{abstract}


\section{Misaligned trust and incentives: The problem of optimization systems}

Fairness solutions emerge as a response to the problems arising from the current widespread application of machine learning. Research on fairness has \emph{pre-eminently} focused on decision-making and recommender systems~\cite{buolamwini2018gender,dwork2012fairness,PleissRWKW17}. Such systems take data from one individual as input and output a decision or a recommendation, based on some algorithm, that has a direct impact on that individual. For example, recidivism prediction systems, which are
well studied in fairness literature as well as news media, take as input the data of a criminal defendant and predict how likely the individual is to reoffend~\cite{perry2013predictive}. Fairness tools act on the algorithm to prevent or minimize inequitable outcomes for different (historically marginalized) groups~\cite{corbett2018measure,chouldechova}.
Over the last years, researchers have proposed various definitions of fairness~\cite{corbett2018measure,PleissRWKW17,Berk18}, and there is a lack of consensus as to what the paradigm encompasses. As such, we discuss fairness in broad terms and focus on the state-of-the-art. We illustrate how the field's current focus limits fairness' ability to mitigate negative outcomes of optimization systems on individuals, populations, and environments.

Recent reports about popular technological service providers invite us to rethink the scope of fairness frameworks and their underlying trust models. 
Waze, the traffic app owned by Google, is frequently reported to disregard the negative impacts of routing vehicles through streets that are inadequate for heavy traffic~\cite{Lop18}. 
A class action lawsuit against Facebook claims that, in order to increase ad revenue, the company inflated the amount of time users viewed videos by as much as 900\%~\cite{Kat18}. 
Due to imbalance in the data used to create their maps, \Pokemon\ Go is known to spawn less \Pokemon in rural areas and majority-black neighborhoods~\cite{pogorace,pogofair}, illustrating how inequality is not limited to employment, income, and housing, but extends to who has leisure time and how it is spent~\cite{Henderson2014}. Despite their attempts to address this biased treatment using traditional means to process their data, Niantic, the company behind the game, has a difficult time fixing the issue~\cite{pogonotfixed}.
These stories highlight a number of problems: the Waze story presents a system that may treat non-users and their environments unfairly; the Facebook story shows that even in the presence of powerful stakeholders and financial incentives, service providers may behave dishonestly or escape recourse by claiming these are machine learning accidents~\cite{Kat18}; and the \Pokemon Go example demonstrates that even when providers are willing, solving fairness problems may require looking beyond possible pre-processing of the input data. 
These examples cast doubt on the ability of service providers to honestly, or effectively, apply fairness solutions to their systems. 


These  are examples of optimization systems: they go beyond simple decision making or recommendation, incorporate real-time feedback from users~\cite{Kaldrack2015}, interact with the environment in which they are deployed, and are optimizing over variables that are constantly changing. 
Optimization systems increasingly rely on machine learning to make operational decisions like the selection of software features, the orchestration of cloud usage, the design of user interaction and growth planning~\cite{GursesVHoboken18}, etc. In contrast to traditional information systems, which treat the world as a static place to be known and focus on storage, processing, transport, and organizing information, optimization systems consider the world as a place to sense and co-create. They seek maximum extraction of economic value by optimizing the capture and manipulation of actions and environments~\cite{Agre1994,CurryPhillips03}. Ride sharing apps rely on optimization to decide on the prices of rides; navigation apps to propose best routes; banks to decide whether to grant a loan; and advertising networks to select the best advertisement to show a user, all driven by an interest in the greatest return on investment. As a consequence,
for the \newterm{optimization system providers} (\deployers), profit maximization and fairness may easily be conflicting goals, i.e., \deployers may lack the necessary incentives to apply fairness solutions honestly or at all.


Furthermore, optimization systems apply a logic of operational control that focuses on outcomes rather than the process ~\cite{Poon2016}. While this introduces efficiency and allows systems to scale, it also poses social risks and harms such as social sorting, mass manipulation, majority dominance, and minority erasure. In summary, these systems create substantial \newterm{externalities} ---
situations in which the actions of a group of agents, e.g., consumption, production, and investment decisions, have ``significant repercussions on agents outside of the group''~\cite{Starrett11}.


Some common externalities intrinsic to optimization systems are i) disregard for non-users and environments which results in non-users being outside the optimization model~\cite{Lop18}; ii) disregard for certain users by providing the most benefit to a subset of ``high-value'' users~\cite{Huf16, Tas16}; iii) externalization of exploration risks to users and environments, in which the systems benefit from experimentation while putting exploration risks associated with environmental unknowns onto users~\cite{BirdBCW16}; iv) distributional shift, in which optimization systems are built on data from a particular domain and flounder when deployed in a different environment~\cite{mcm11,fie04,sug17}; v) unfair distribution of errors, in which algorithms learn to maximize success by favoring the most likely option, e.g., misclassify minorities while maintaining high accuracy~\cite{buolamwini2018gender,Har14}; vi) promotion of unintended actions to fulfill intended outcomes, where systems find shortcuts to their optimization goals, also known as ``reward hacking''~\cite{AmodeiOSCSM16,cheatingBots}; and vii) mass data collection, where, in the pursuit of more accurate inferences, resources and power are concentrated to a few data holders ~\cite{Poon2016}.

Such externalities that arise in optimization systems are beyond unfairness due to the inputs and outputs of the algorithms, or may not be caused by algorithmic unfairness at all. That is, an ``unbiased'' algorithm can still have unfair consequences or externalities. Such problems cannot be solved by diligent, and less so by dishonest, service providers. Instead, they require additional models and techniques to reason about strategies to counter them.


\section{Solutions by design?}

We argue that \deployers may not have the incentives or means to apply fairness solutions to their systems and that such solutions may not be sufficient to address the many externalities of optimization systems. We discuss in greater detail why fairness, or more broadly solutions by design, may fall short of the task.

Existing framings of externalities do not capture conflicts between the goals of the system and fairness. Typically, AI safety experts argue that the harms of optimization systems arise because \deployers \emph{``choose `wrong' objective functions''} or \emph{``lack sufficient good-quality data''}, i.e., flaws and accidents amount to \emph{poor design}~\cite{AmodeiOSCSM16,Kat18}. Fairness frameworks often follow suit.
One response, therefore, is to devise countermeasures that allow \deployers to prevent or minimize the occurrence of these \emph{`flaws'}~\cite{AmodeiOSCSM16}, with the underlying assumption that, given adequate tools and means, \deployers will strive to \emph{`fix'} or \emph{`correct'} their systems. While developing methods that can improve the design of optimization systems is absolutely necessary, this alone is insufficient to counter optimization systems' negative externalities. 
The \emph{accidents} framing dismisses the possibility that those design choices may in fact be intentional, i.e., that the objective functions underlying optimization systems may actively \emph{aspire for asocial or negative environmental outcomes}. \deployers may lack incentives to maximize society's welfare as opposed to their own benefit and, therefore, avoid applying fairness frameworks or demonstrate adversarial behavior.

Moreover, when an objective function aspires for asocial outcomes, applying fairness may come to exacerbate problems. \emph{Decontextualized from the goal of the system}, fairness solutions may ensure that subgroups of the population are equally affected by the algorithm, only considering the distribution of outcomes as they are aligned with the goal of the algorithm. In other words, failing to question the system's utility may not consider whether the objective function itself is just, ensuring
only that people are equally subject to its effects~\cite{dressel2018accuracy,binns2017fairness}. For instance, a predictive policing application that is constrained for fairness but does not take into account the negative effects that predictive policing has on vulnerable populations, or a credit scoring algorithm that is tuned to ensure that sub-prime loans are fairly distributed, are likely to lead to unjust outcomes. A notable exception to this frame in fairness literature is written in
\citet{Corbett-Davies:2017} which considers the impact on utility in the form of a measure of public safety. Decontextualization can further intensify negative outcomes when \deployers pick fairness frameworks that limit their engagement to legal conceptions of protected identities, skirting more complex problems that cannot be reduced to ``identity attributes'' or the ``cause-and-effect'' model inherent to legal definitions~\cite{Hoffmann2018}. Such models may be depicted as fair while avoiding responding to unjust outcomes. 

The asociality of the objective function may be concealed by \emph{blindness to allocation of resources}. The quality of an allocation can be determined by different qualities, including, but not limited to, fairness. For example, one could consider that the sum of all payoffs should be as high as possible, or that the worst-off agents in a system should be as well-off as possible. One could also consider injustice in greater time frames and use models that engage in restorative justice~\cite{Schock2018}. While fairness will aspire for parity in allocation of resources, it does not always consider these other qualities. For example, fairness guarantees may be applied to the rider-driver matching algorithms with the intention to ensure that Uber drivers are equally likely to find riders, access surges, and earn similar wages regardless of their belonging to a protected class. This does not, however, deal with the fact that Uber optimizes the matching to maximize profit while minimizing driver wages. In such an instance, distributing suppressed wages fairly is not quite the ultimate objective.


Even when \deployers have incentives to address relevant problems optimization causes through fairness, they may not be in a position to do so. \emph{\deployers may lack knowledge about the needs of those affected by optimization}, even when they strive to collect the necessary data to mend optimization outcomes: such (environmental) data may simply not be available for capture. In practice, data often represents what is easy to capture and thus provides a biased account of the people and
environments it supposedly measures, leaving out key nuances required to improve optimization~\cite{greenfield2017radical}. 



Fairness literature often considers decision-making as static, i.e., the algorithm and the environment in which it is deployed are fixed. However, under optimization systems, the decision-making model does change, e.g., predictive policing results in increased crime reporting for patrolled neighborhoods if the algorithm is not updated to account for the fact that police visit certain neighborhoods more often than others~\cite{EnsignFNSV17}. While exceptions exist in which future changes of environments are considered, e.g., the work on fairness in reinforcement learning~\cite{Jabbari16}, in general fairness disregards problems that go beyond discrimination in algorithmic decision making and is therefore \emph{ill-equipped to prevent the negative effects that arise when agents evolve}~\cite{LiuDRSH18}.

Fairness also often stays limited to analyzing an algorithm's inputs and outputs, but in optimization systems many externalities occur only after the system is introduced into an environment. Because the state that fairness considers often does not reflect changes in the environment, fairness solutions are made \emph{unaware of and cannot account for most externalities that surface post-deployment}. For example, Waze may provide its users with \emph{optimal} routes in a way that satisfies a given notion of fairness. However, fairness cannot reason about the effect that the \emph{fair} routes have on the environments that those routes traverse, i.e., congestion may increase on surface roads. This impact, especially when multiple routing applications are at play, may become evident only after deployment.
Moreover, because fairness solutions are intended to be deployed by \deployers, they are often limited to protecting only their users, while disregarding the non-users. Waze, for example, has no explicit way of getting feedback from people who are affected by its routing decisions, yet do not use the application.

We argue that \deployers should internalize potential harms and risks, e.g., through stringent design practices, regulation and taxation, and democratic forms of governance. Short of this, we can consider \deployers as potentially unable, lacking incentives, or unwilling to address the externalities of their optimization systems, rather than considering these as the result of poor design choices or accidents. 

\section{Rethinking the trust model: solutions from the outside}

The points above indicate that fairness, conceptually and with respect to its assumptions about the trust and ability of \deployers, is limited in its capacity to respond to externalities of optimization systems. 
We need new mental models and techniques that enable designers to reason about strategies that not only counter the negative effects of optimization from \emph{within} the system, but also from \emph{outside} of the system. Furthermore, we may need to capture alternative optimization functions that are different from those embedded in the service provider's own optimization algorithms, to the extent that they may consider variables and contour conditions not even present in the original optimization system.

We propose that approaches to mitigate the externalities of optimization systems should build on alternative trust models that do not rely solely on \deployers and should consider factors of the environment not explicitly present in the machine learning models. A number of existing works have considering hedging OSPs against users ``gaming'' the optimization system in their favor~\cite{bruckner2011stackelberg,hardt2016strategic}. Protecting the optimization system or striving for an equilibrium between adversarial users and the system's goals, however, have been shown to reinforce inequality~\cite{hu2018disparate} and disproportionately benefit the \deployers~\cite{milli2018social}.
 
With this in mind we invite the community to think about how to design and implement a new class of defenses to enable those affected by optimization systems to influence, alter, and counter these systems \emph{from the outside}. We call these defenses Protective Optimization Technologies (POTs). POTs respond to the discontents of machine learning that fairness frameworks cannot completely address. They are intended to empower people and environments affected by optimization systems to intervene when \deployers fail to respond to their needs. They rely on explicit modeling and evaluation of the impact of optimization systems on populations and environments, broadening the scope of unfairness and mitigation techniques.

POTs development requires to analyze how events (or lack thereof) affect users and environments, and then find the means to reconfigure these events, i.e., influence the system's outcomes, by poisoning the training data thus modifying the optimization constraints, or crafting alternative system inputs to counteract the optimization effect. We specifically conceive POTs to address the negative externalities of optimization, and as such it is crucial to take a holistic perspective, considering the interaction of the algorithm with the rest of the optimization system and the environment. These ideas are inspired by people's strategies to counter the negative effects of optimization systems by manipulating inputs to the system and not changing the algorithms themselves, in order to achieve a desired, more balanced, output. For instance, neighborhood dwellers negatively affected by Waze's routing have fought back by reporting road closures and heavy traffic on their streets --- to have Waze redirect users out of their neighborhoods. Uber drivers have colluded to induce \emph{surge} pricing and temporarily increase their revenue by simultaneously turning off their apps, inducing surge, and turning the app back on to take advantage of the increased pricing in the area~\cite{Mhlmann2017HandsOT}. Adnauseam, a browser plugin seeks to pollute advertisting systems' profiling by clicking on random ads in the background in order to render inferred user profiles useless.



These ad-hoc examples demonstrate that it is possible to counter optimization externalities from the outside. As such, their principles can inspire the design of more formal POTs. For instance, these could be based on adversarial machine learning techniques used to bias the optimization system responses to reduce their negative impact on users and environment. Such an idea appears in recent calls to develop POTs for civil liberties~\cite{Kumar18}, and is already prevalent in Privacy-Enhancing Technologies (PETs) literature~\cite{McDonaldACS12, CaiZJJ12, che17} --- in the spirit of which we attend to the optimization problem. 

In this position paper we are just scratching the surface of the problem and the possible solution space. We hope that this inspires the community to engage in a deep discussion about the assumptions underlying current solutions. We consider this an indispensable step before we can build effective robust solutions to address the negative effects of machine learning, and more broadly optimization systems.

\section*{Acknowledgements}

This work was supported in part by the Research Council KU Leuven: C16/15/058, and the European Commission through KU Leuven BOF OT/13/070, H2020-DS-2014-653497 PANORAMIX, and H2020-ICT-2015-688722 NEXTLEAP. Seda Gürses is supported by a Research Foundation - Flanders (FWO) Fellowship. 

\bibliographystyle{unsrtnat}
\bibliography{main}

\end{document}